\documentclass[aps,prl,showpacs,floats,twocolumn,floats,superscriptaddress,floatfix]{revtex4}
\usepackage{bm}
\usepackage{times}
\usepackage{verbatim}
\usepackage{graphicx}
\usepackage{graphics,epsfig}
\usepackage{theorem}
\usepackage{makeidx}
\usepackage{amsmath}
\usepackage{epic}
\usepackage{amscd}
\usepackage{bbm}
\usepackage{xy}
\usepackage{amssymb}
\usepackage{epsfig}

\begin{document}
\preprint{LA-UR 08-2024}
\newif\iffigs 
\figstrue
\iffigs \fi
\def\drawing #1 #2 #3 {
\begin{center}
\setlength{\unitlength}{1mm}
\begin{picture}(#1,#2)(0,0)
\put(0,0){\framebox(#1,#2){#3}}
\end{picture}
\end{center} }

\newcommand{\ul}{{\bm u}_{\scriptscriptstyle \mathrm{L}}}
\newcommand{\deltal}{{\delta}_{\scriptscriptstyle \mathrm{L}}}
\newcommand{\ue}{\mathrm{e}}
\newcommand{\ui}{\mathrm{i}\,}
\newcommand{\kg}{{k_{\scriptscriptstyle \mathrm{G}}}}
\def\v{\bm v}
\def\x{\bm x}
\def\k{\bm k}
\def\ds{\displaystyle}

\title{Hyperviscosity, Galerkin truncation and
bottlenecks in turbulence} 
\author{Uriel Frisch}
\affiliation{Labor. Cassiop\'ee, OCA, UNS, CNRS, BP 4229, 06304 Nice
  cedex 4, France}
\author{Susan Kurien}
\affiliation{CNLS \& Theoretical Division, Los Alamos National Laboratory, 
Los Alamos, NM 87545, USA}
\author{Rahul Pandit}
\affiliation{Centre for Condensed Matter Theory, 
Department of Physics, Indian Institute of Science, 
Bangalore, India}
\author{Walter Pauls}
\affiliation{Max Planck Institute for Dynamics and Self-Organization, 
Goettingen, Germany}
\author{Samriddhi Sankar Ray}
\affiliation{Centre for Condensed Matter Theory, 
Department of Physics, Indian Institute of Science, 
Bangalore, India}
\author{Achim Wirth}
\affiliation{LEGI, CNRS, Grenoble, France}
\author{Jian-Zhou Zhu}
\affiliation{CNLS \& Theoretical Division, Los Alamos National Laboratory, 
Los Alamos, NM 87545, USA}

%
%

\begin{abstract}

It is shown that the use of a high power $\alpha$ of the Laplacian in the
dissipative term of hydrodynamical equations leads asymptotically to truncated
inviscid \textit{conservative} dynamics with a finite range of spatial Fourier
modes. Those at large wavenumbers thermalize, whereas modes at small
wavenumbers obey ordinary viscous dynamics [C.~Cichowlas {\it et al.} Phys.\
Rev.\ Lett.\ {\bf 95}, 264502 (2005)]. The energy bottleneck observed for
finite $\alpha$ may be interpreted as incomplete thermalization.  Artifacts
arising from models with $\alpha > 1$ are discussed.


\end{abstract}


\date{\today}

\maketitle

%
%
A single Maxwell daemon embedded in a turbulent flow
would hardly notice that the fluid is not exactly in
thermal equilibrium because incompressible turbulence, even at very
high Reynolds numbers, constitutes a tiny
perturbation on thermal molecular motion. Dissipation in real fluids is just
the transfer of macroscopically organized (hydrodynamic) energy
to molecular thermal energy. 
Artificial microscopic
systems can act just like the real one as far as the
emergence of hydrodynamics is concerned; for
instance, in lattice gases the ``molecules'' are discrete
Boolean entities~\cite{FHP} and thermalization is easily observed at high 
wavenumbers~\cite{SSB88}. Another example has been
found recently by Cichowlas {\it et al.}~\cite{Cichowlas}
wherein the Euler equations of ideal non-dissipative
flow are (Galerkin) truncated by keeping only a finite
-- but large -- number of spatial Fourier harmonics.
The modes with the highest wavenumbers $k$ then
rapidly thermalize through a mechanism discovered by
T.D.~Lee~\cite{Lee52} and studied further by R.H.~Kraichnan \cite{KrAbEq}, leading in three dimensions (3D)
to an equipartition energy spectrum $\propto k^2$.
The thermalized modes act as a fictitious microworld on
modes with smaller wavenumbers in such a way that the
usual dissipative Navier--Stokes dynamics is
recovered at large scales \footnote{A similar mechanism, involving
sound waves, explains why superfluid and ordinary turbulence can be
similar; see C.~Nore, M.~Abid,
and M.E.~Brachet, Phys.\ Rev.\ Lett. {\bf 78}, 3896
(1997).}.

All the known systems presenting thermalization are \textit{conservative}.  As
we shall show themalization may be present in \textit{dissipative}
hydrodynamic systems when the dissipation rate increases so fast with the
wavenumber that it mimics ideal hydrodynamics with a Galerkin
truncation. This is best understood by considering hydrodynamics with
\textit{hyperviscosity}: the usual momentum diffusion operator (a Laplacian)
is replaced by the $\alpha$th power of the Laplacian, where $\alpha>1$ is the
dissipativity. Hyperviscosity is frequently used in turbulence modeling to
avoid wasting numerical resolution by reducing the range of scales over which
dissipation is effective~\cite{hyperviscosity}.

The unforced hyperviscous 1D Burgers and multi-dimensional  
incompressible  Navier--Stokes (NS) equations are:
\begin{equation}
\partial_t v +v\partial_x v =-\mu \kg^{-2\alpha}
(-\partial_x ^2)^{\alpha}v ;
\label{hyperburgers}
\end{equation}
\begin{equation}
\partial_t \v+\v\cdot \nabla \v =-\nabla p -\mu
  \kg^{-2\alpha}(-\nabla ^2)^{\alpha}\v;\quad
  \nabla\cdot \v=0\;.
\label{hyperns}
\end{equation}
The equations must be  supplemented with suitable initial and
boundary conditions. We employ $2\pi$-periodic
boundary conditions in space,  so that we can use Fourier
decompositions such as $\v(\x) =\sum_{\k} \hat\v_{\k}
\,\ue ^ {\ui \k\cdot \x}$. Note that minus the
Laplacian is a positive operator, with Fourier
transform $k^2$, which can be raised to an arbitrary power $\alpha$.
The coefficient $\mu$ is taken positive to make the hyperviscous
operator dissipative. The \textit{Galerkin wavenumber} 
$\kg>0$ is chosen \textit{off-lattice}
so that no wavenumber is exactly equal to $\kg$.  In
Fourier space the hyperdissipation rate is
$\mu(k/\kg)^{2\alpha}$, where $k \equiv
|\k|$.

If we now hold $\mu$  and $\kg$ fixed and let $\alpha \to \infty$ we see
that the hyperdissipation rate tends to zero, for $k< \kg$, and to infinity, for
$k>\kg$. This implies that \textit{in the limit of
infinite dissipativity, the solution of a hyperviscous
hydrodynamical equation converges to that of the
corresponding inviscid equations Galerkin-truncated at
wavenumber} $\kg$.

%
%

To define inviscid Galerkin truncation precisely, we
rewrite Eqs.~(\ref{hyperburgers}) and (\ref{hyperns}) in
the abstract form $\partial_t v = B(v,v) +L_\alpha v$,
where $B$ is a quadratic form representing the
nonlinear term (including the pressure $p$ in the NS
case). The truncation projector $P_\kg$ is the linear,
low-pass filtering operator that, when applied to $v$,
sets all Fourier harmonics with $k> \kg$ to zero. The inviscid,
Galerkin-truncated equation, with initial condition
$v_0$, is
\begin{equation}
\partial_t u = P_\kg B(u,u), \qquad u_0=P_\kg v_0\;. 
\label{igt}
\end{equation}
Since $u$ can be written in terms of a finite number
of modes with $k <\kg$, Eq.~\eqref{igt} is a dynamical
system of finite dimension. In addition to momentum,
it conserves the energy and other quadratic
invariants for the inviscid equations~\cite{KrAbEq}.
There is good numerical evidence -- but no
rigorous proof --
that the solutions of the Galerkin-truncated inviscid
Burgers and 3D Euler equations tend, at large times,
to statistical equilibria defined by their respective
invariants. 


%
%

A rigorous proof of the convergence, as $\alpha \to \infty$, of solutions of the
hyperviscous Burgers equation \eqref{hyperburgers}
and of the hyperviscous NS equation \eqref{hyperns} in any dimension to those of
the associated Galerkin-truncated, inviscid equation
will be given elsewhere. It uses standard tools of
functional analysis; 
note that the formidable mathematical difficulties
that beset the ordinary ($\alpha =1$) 3D NS equation 
disappear for $\alpha\ge 5/4$ \cite{lions}. 

From a physicist's
point of view the convergence result looks rather obvious, though
it has hardly been noted before (see, however, Refs.~\cite{KrPR58,JJ94,boyd94}):
as $\alpha \to \infty$ all the modes with $k>\kg$ are
immediately suppressed by an infinite dissipation,
whereas those with $k<\kg$ obey inviscid truncated
dynamics.  
 Not surprisingly, the fate of couplings between
   triads of modes whose wavenumbers straddle $\kg$ is a delicate point.
In a Galerkin 
truncation any such triad should be left out. It may be shown 
that for $\alpha \to \infty$ such straddling couplings are suppressed, not
only for the Burgers  and NS equation  but also for the hyperviscous magnetohydrodynamical
equations  and for some turbulence
closures, specifically,  the Direct Interaction Approximation
(DIA)~\cite{KrPR58} and the
Eddy-Damped-Quasi-Normal-Markovian (EDQNM)
approximation~\cite{EDQNM}.  Hence the convergence to the corresponding
Galerkin-truncated equations holds for all the aforementioned equations
in any dimension of space.

There are, however, interesting exceptions among hydrodynamical equations
for which the result does not hold. They include the kinetic theory of resonant wave
interactions~\cite{ZLF} and the Markovian Random Coupling Model
\cite{MRCM}. Indeed, the resonant wave interaction 
theory arises in the limit when the period of the waves goes to zero and
this limit does not commute with the limit of a vanishing damping time
for modes having $k>\kg$; a similar remark can be made about the MRCM
equation.

Let us stress that systems with a \textit{finite}
dissipativity -- however large -- are quite
different from Galerkin-truncated systems. For
example, consider the 3D NS equation with a random
force, delta-correlated in time, for which we know
the mean energy input $\varepsilon$ per unit volume.
It is still true that, for $\alpha \to \infty$, the
solution of this equation converges to that of the
Galerkin-truncated equation, but this time with a
random force. If $E_0$ is the initial energy, this
solution has a mean energy $E(t)=E_0+\varepsilon t$,
which grows indefinitely in time. But, as soon as $\alpha$ is given a
finite value, however large, a statistical steady state,
in which energy input and hyperviscous energy
dissipation balance, is achieved at large times. Such
a steady state presents an interesting interplay of
thermalization and dissipation, when $\alpha$ is
large, as we show below.

%
%

\begin{figure}
 \iffigs 
 \centerline{%
 \includegraphics[scale=0.28]{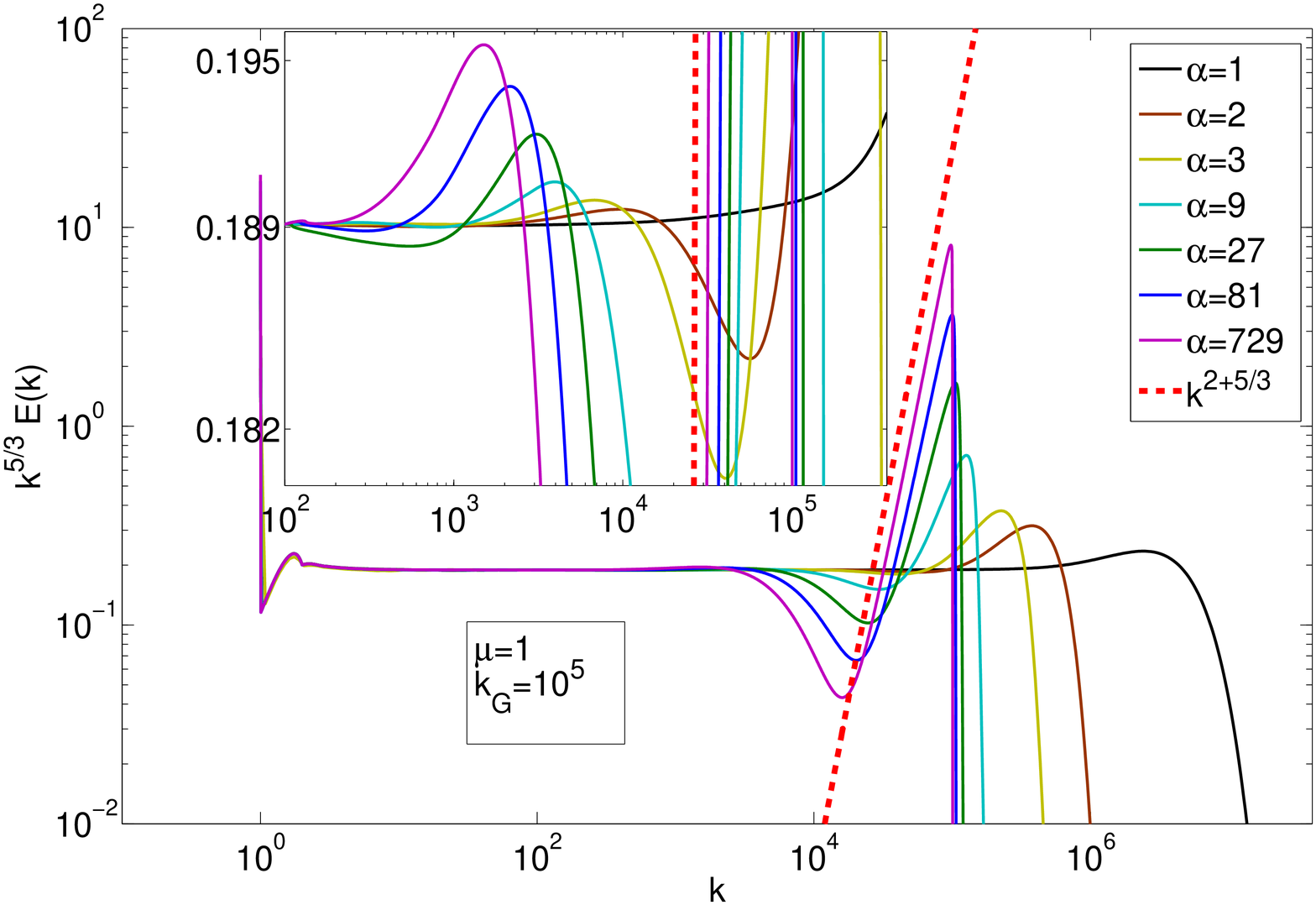}%
}
 \else\drawing 65 10 {Jian-Zhou Zhu many $\alpha$}
 \fi 
\caption{(Color online) Log-log plots of the
compensated spectrum $k^{5/3}E(k)$ versus $k$ from a numerical 
integration of the hyperviscous EDQNM Eq.~\eqref{EDQNM} 
for different values of $\alpha$; inset:
enlarged spectra showing a secondary bottleneck (see text).}
\label{f:edqnm}
\end{figure}

The direct numerical simulation (DNS) of the
Galerkin-truncated 3D Euler equations in
Ref.~\cite{Cichowlas} used $1600^3$ Fourier modes.
Large-$\alpha$ simulations of Eq.~(\ref{hyperns}) would
require significantly higher resolution to identify
the various spectral ranges that we can expect,
namely, inertial, thermalized, and far-dissipation
ranges and transition regimes between these.
Fortunately, Bos and Bertoglio~\cite{BB} have shown
that key features of the Galerkin-truncated Euler
equations, such as the presence of inertial and thermalized ranges, can be reproduced by the two-point
EDQNM closure~\cite{EDQNM} for the energy spectrum.
For Eq.~(\ref{hyperns}), with stochastic, white-in-time,
homogeneous and isotropic forcing with spectrum
$F(k)$, the  hyperviscous EDQNM equations are
\begin{equation}
\left.
\begin{array}{l}
\ds \left(\partial_t +2\mu \left(\frac{k}{\kg}\right)^{2\alpha}\right)
E(k,t)  
=
\int\!\!\!\!\int_{\triangle_k}dpdq\,\theta_{kpq}\,\times \\[1.6ex]
\ds b(k,p,q)\frac{k}{pq}E(q,t)\left[k^2E(p,t)-p^2E(k,t)\right]+F(k)\;,\\[1.6ex]
\ds \theta_{kpq}=\frac{1}
{\mu_k+\mu_p+\mu_q}\;, \quad b(k,p,q)=\frac{p}{k}(xy+z^3)\;,\\[1.6ex]
\ds \mu_k = \mu \left(\frac{k}{\kg}\right)^{2\alpha} +\lambda
\left[\int_0^k p^2E(p,t)dp\right]^{\frac{1}{2}}\;.\\[1.6ex]
\end{array}
\right\}
\label{EDQNM}
\end{equation}
Here $E(k,t)$ is the energy spectrum, $\triangle_k$ defines the set
of $p\ge 0$ and $q\ge 0$ such that $k$, $p$, $q$ can form 
a triangle, 
$x$, $y$, $z$ are the cosines of its angles and
the eddy-damping parameter $\lambda$ is expressed in terms of the
Kolmogorov constant. The EDQNM equations have been
studied numerically for more than three decades~\cite{HerringEDQNM}, 
but their hyperviscous
versions Eq.~(\ref{EDQNM}) present new difficulties that we
overcome as follows.  Since we are interested in the
steady state we use an iterative method: the emission
term, $E(p,t)E(q,t)$ in Eq.~(\ref{EDQNM}), is considered
as a renormalization of the force $F(k)$; the
absorption term, $E(q,t)E(k,t)$, is treated as a
renormalization of the hyperviscous
damping~\cite{KrPoF64}.  We then construct a sequence
of energy spectra that, at stage $n+1$, is just the
renormalized force divided by the renormalized
damping, both based on stage $n$.  This gives rapid
convergence to the steady state 
at low wavenumbers; but, beyond a certain ($\alpha$-dependent)
wavenumber, convergence slows down dramatically and it is better to
use time marching to obtain the steady state.  At large values of $k$
and $\alpha$ the problem becomes very stiff, so we use a slaved
fourth-order Runge--Kutta scheme~\cite{CM02}.  We discretize $k$
logarithmically, with $N_c$ collocation points per octave. Triad
interactions involving wavenumber ratios significantly larger than
$N_c$ are poorly represented~\cite{LS78}; so, since wavenumber ratios
of up to 50 play an important role for large $\alpha$, we have used
$N_c \lesssim 90$; this is computationally demanding because the
complexity of the code is $O(N_c^3)$.  We force at the lowest
wavenumber ($k=1$) in our numerical study of Eq.~(\ref{EDQNM}) with
$\kg =10^5$, $\lambda =0.36$, and $1 \leq \alpha \leq 729$. The
resulting compensated, steady-state energy spectra $k^{+5/3}E(k)$ are
shown in Fig.~\ref{f:edqnm}; flat regions, extending over two to five
decades of $k$ (depending on $\alpha$), are close to the Kolmogorov
inertial range; for large $\alpha$ there is a distinct thermalized
range with $E(k) \sim k^2$ (also found in the transition between
classical and quantum superfluid turbulence~\cite{LNR}), as we expect
from our discussion of the Galerkin-truncated Euler
equations~\footnote{A scaling argument suggests that the width of the
thermalized range grows as $\alpha ^{1/4}$ (up to logarithms);
checking this numerically at high $\alpha$ is difficult because there
is a boundary layer near $\kg$, of width $O(\kg/\alpha)$, in which a
transition occurs from very small to very large dissipation.}. In the
far-dissipation range $k > \kg$ the spectra fall off very rapidly
\footnote{Dominant balance gives for the leading term 
a decreasing exponential with a $k^{2\alpha +1}$ prefactor for the 3D
EDQNM  case and a $k^{2\alpha -2}$ prefactor for the 1D Burgers case.}.
For all values of $\alpha$ the far-dissipation range is
preceded by a bump or \textit{bottleneck}.  It is also observed, in
some experiments \cite{bottleexper} and DNS of Navier--Stokes, with a shape that is quite independent
of the Reynolds number \cite{bottlesimul}.  The bottleneck for $\alpha
= 1$ has previously been explained as the inhibition of the energy
cascade from low to high wavenumbers because of viscous suppression of
the cascade in the dissipation range~\cite{falkov}. Our work provides
an alternative explanation: the usual bottleneck may be viewed as
\textit{incomplete thermalization}.


At large values of $\alpha$ the thermalized range
gives rise to an eddy viscosity $\nu_{\rm eddy}$. This
acts on modes with wavenumbers lower than those in the
thermalized range; the corresponding damping rate is
$\nu_{\rm eddy} k^2$.  The eddy viscosity can be
expressed as an integral over the thermalized
range~\cite{BB,LS78}. 
As $\alpha$ grows, so does $\nu_{\rm eddy}$ and,
eventually, the renormalized viscous damping
overwhelms the hyperviscous damping for modes at low
wavenumbers (below those in the thermalized range).
The dynamics of these modes is then governed by the
usual $\alpha =1$ equation. Not surprisingly, then,
we find a pseudo-dissipation range around $k \simeq
10^4$ that is shown in an expanded scale in the inset
of Fig.~\ref{f:edqnm}; a similar range for the
Galerkin-truncated case is discussed in Ref.~\cite{BB}
and is already visible in the DNS of
Ref.~\cite{Cichowlas}.  For large $\alpha$ the inset
of Fig.~\ref{f:edqnm} also shows a \textit{secondary
bottleneck} range for $10^3 < k < 10^4$; this may be viewed
as the usual ($\alpha =1$) EDQNM bottleneck stemming from
$\nu_{\rm eddy}$ \footnote{The secondary bottleneck overshoots by
$\simeq 3-5\%$, compared to $\simeq 20\%$ for the usual bottleneck, 
perhaps because of higher-order terms in the renormalized
damping, which are beyond the eddy-viscosity
approximation.}. 



%
%

Our results apply to compressible flows also. We have studied the 
simplest instance, that is the unforced hyperviscous 1D Burgers
equation \eqref{hyperburgers} \footnote{It is easily shown that the ordinary
  Burgers equation produces no bottleneck.}. Its
solution converges to the \textit{entropy}
solution, i.e., the standard solution with shocks,
obtained when $\kg \to \infty$ for any $\alpha\ge1$ 
\cite{ET04}. Here we are interested in the
large-$\alpha$ behavior at fixed $\kg$. 
We do not have to resort to closure now since we can
solve the primitive equation \eqref{hyperburgers} directly
by a pseudospectral method. If we choose a single
initial condition the resulting spectrum is noisy because, 
unlike the ordinary Burgers equation, its Galerkin-truncated 
version and thus also the high-$\alpha$ versions are believed to be chaotic dynamical
systems~\cite{Majda}. 
So we solve
\eqref{hyperburgers} with the two-mode random initial condition 
$v_0(x) = \sin x +\sin (2x+\phi)$, where $\phi$ is 
distributed uniformly in
the interval $[-\pi,\, \pi]$. 
We use $2^{14}$ collocation
points  and set $\mu=1$, 
$\kg =  342.1$ and $\alpha =1000$.
\begin{figure}[!h]
 \iffigs 
 \centerline{%
 \includegraphics[scale=0.25]{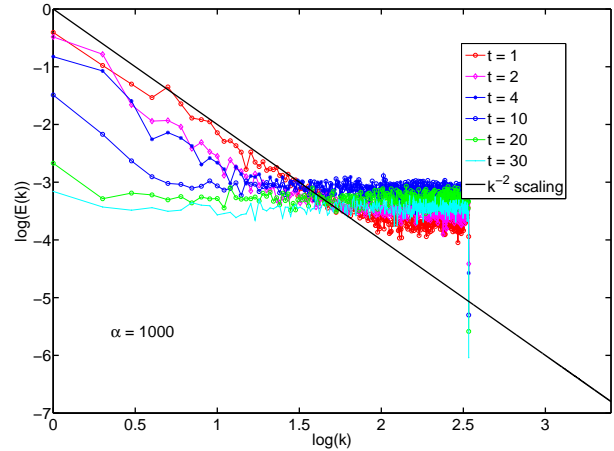}
 }
 \else\drawing 65 10 {SSR's labelled spectra}
 \fi 
\caption{(Color online) Log-log plots of averaged
energy spectra (see text) versus wavenumber $k$ for the 
hyperviscous Burgers 
equation \eqref{hyperburgers} at various times.}
\label{f:burgers-spectra}
\end{figure}
In Fig.~\ref{f:burgers-spectra} we show the Burgers
energy spectrum $E(k) = \mid \tilde{v}(k)\mid^2$,
averaged over 20 realizations of the phase $\phi$ at
various times.  At the latest output times the
spectrum is almost completely flat, i.e., thermalized,
with equipartition of the energy between all the
Fourier modes.
At earlier times $E(k)$ behaves approximately as $k^{-2}$
in an inertial range that corresponds to shocks in
physical space; there is a thermalized range at higher
wavenumbers up to $\kg$; for $k > \kg$ the spectrum
falls  very rapidly.  No pseudo-dissipation
range is observed here between the inertial and
thermalized ranges as seen in the 3D NS case
(Fig.~\ref{f:edqnm}). Perhaps  the
data are too noisy, but a careful examination of
$v(x)$ in physical space indicates that this
phenomenon might arise from the compressible nature of
the Burgers dynamics: thermalization  begins over the
whole physical range (as high-frequency noise with
wavenumber $\approx \kg$); noise generated close to
shocks is absorbed by them and not
enough is left to produce any appreciable eddy
viscosity that could broaden the shocks.  
We now summarize our main findings from the study of hyperviscous
hydrodynamical equations  with powers $\alpha$ of the Laplacian ranging
from unity to very large values.

The simplest results are obtained for very large $\alpha$. The solutions of
the 1D Burgers equation or the Navier--Stokes equations in any space
dimension
$d$ are then very close to the solutions of the corresponding
Galerkin-truncated equations, displaying thermalization at wavenumbers below
$\kg$. The detailed scenario will of course be affected by the dimension
of space. In 3D, with enough resolution, we may be able to observe
up to  five ranges: an inertial range, a secondary bottleneck, a
pseudo-dissipation range, a thermalized range, and a far dissipation
range. Because of enstrophy conservation and of the predominance of
Fourier-space nonlocal interactions, the 2D case is rather special and
deserves a separate study\footnote{In the  2D case several aspects other
than thermalization can be captured by the linear hyperviscous
theory of Ref.~\cite{JJ94}.}.

The most relevant case is of course that of ordinary dissipation ($\alpha
=1$).  The energy-spectrum bottleneck generally observed at high Reynolds
numbers in 3D incompressible turbulence may be viewed as an incomplete
thermalization: as we increase $\alpha$ larger and larger bottlenecks are
present, eventually displaying thermalization on their rising side.

We finally deal with the case of moderately large $\alpha$ of the sort
used in many simulations \cite{hyperviscosity}. How safe is this
procedure and what kind of artifacts can we expect?

Using large values of $\alpha$ in simulations to ``avoid wasting 
resolution'' is hardly advocated by anybody, but we now  understand
what goes wrong: a huge thermalized bottleneck will develop at high
wavenumbers, whose action on smaller wavenumbers  is an ordinary $\alpha
=1$ dissipation with an eddy viscosity much larger than what would be
permissible in a normal $\alpha =1$ simulation.

When $\alpha$ is chosen just a bit larger than unity (e.g. $\alpha =2$ which
is standard in oceanography \cite{hyperviscosity}) the advantage of
widening the inertial range may be offset by artifacts at
bottleneck scales; indeed, even an incomplete thermalization will bring the
statistical properties  of such scales closer to Gaussian,  thereby reducing
 the rather strong intermittency which would otherwise be expected
\footnote{For $\alpha =1$ a lull in the growth of intermittency
  at bottleneck scales may already be observed; cf. Fig.~2 Panel~R4 of
  J.-Z.~Zhu, Chin.\ Phys.\ Lett. {\bf 8}, 2139 (2006).}. For similar reasons
spurious isotropization can be expected for  problems with an
anisotropic constraint, such as rapidly rotating or
stratified flow or MHD with a strong uniform magnetic
field.

We thank J.~Bec, R.~Benzi, M.E~Brachet, W.~Bos, J.R.~Herring, D.~Mitra,
K.~Khanin, A.~Majda, S.~Nazarenko,  D.~Schertzer, E.S.~Titi and V.~Yakhot
 for useful discussions. JZZ thanks the
CNLS fellows and visitors for many interactions.
UF and WP were partially supported by
ANR ``OTARIE'' BLAN07-2\_183172.  RP and SSR thank DST, JNCASR,
and UGC (India) for support and SERC (IISc) for
computational resources. UF, RP, and WP are part of
the International Collaboration for Turbulence
Research (ICTR).  SK and JZZ were
funded by the Center for Nonlinear Studies LDRD
program and the DOE Office of Science Advanced
Scientific Computing Research (ASCR) Program in
Applied Mathematics Research.



\end{document}